\documentclass[runningheads,a4paper]{llncs}
\usepackage{amssymb}
\usepackage{amsmath}
\usepackage{algorithm}
\usepackage{algorithmic}

\usepackage[utf8x]{inputenc}
\usepackage{graphicx}
\usepackage{multirow}
\usepackage{url}
\usepackage{color}
\usepackage{subfigure}

\usepackage{enumitem}
\setlist{
  listparindent=\parindent,
  parsep=0pt,
}

\begin{document}

\authorrunning{Nieves R. Brisaboa et al.}
\titlerunning{Efficient Rep. of Multidimensional Data over Hierarchical Domains}

\title{{Efficient Representation of Multidimensional Data
    over Hierarchical Domains}
\thanks{\footnotesize{Founded in part by Fondecyt 1-140796 (for Gonzalo Navarro);
  and, for the Spanish group, by  MINECO (PGE and FEDER) [TIN2013-46238-C4-3-R];
  CDTI, AGI, MINECO [IDI-20141259/ITC-20151305/ITC-20151247]; ICT COST Action IC1302;
  and by Xunta de Galicia (co-founded with FEDER) [GRC2013/053]. This article was
  elaborated in the context of BIRDS, a European project that has received funding
  from the European Union's Horizon 2020 research and innovation programme under the Marie
  Sklodowska-Curie GA No 690941. } }
  }

\author{Nieves R. Brisaboa \inst{1} \and Ana Cerdeira-Pena\inst{1} \and Narciso L\'opez-L\'opez\inst{1} \and  \\
 Gonzalo Navarro \inst{2} \and Miguel R. Penabad \inst{1} \and Fernando Silva-Coira \inst{1}}
\institute{
           Database Lab., University of A Coru\~na, Spain.
       \email{\{brisaboa,acerdeira,narciso.lopez,penabad,fernando.silva\}@udc.es}
     \and
           Dept. of Computer Science, University of Chile, Chile.
       \email{gnavarro@dcc.uchile.cl}
}
\maketitle

\begin{abstract}

We consider the problem of representing multidimensional data where the
domain of each dimension is organized hierarchically,
and the queries require summary information at a different node in the
hierarchy of each dimension. This is the typical case of OLAP databases.
A basic approach is to represent each hierarchy
as a one-dimensional line and recast the queries as multidimensional range
queries. This approach can be implemented compactly by generalizing to more
dimensions the $k^2$-treap, a compact representation of
two-dimensional points that allows for efficient summarization queries along
generic ranges. Instead, we propose a more flexible generalization, which
instead of a generic quadtree-like partition of the space, follows the
domain hierarchies across each dimension to organize the partitioning. The
resulting structure is much more efficient than a generic multidimensional
structure, since queries are resolved by aggregating much fewer nodes of the
tree.

\end{abstract}

\section{Introduction}

In many application domains the data is organized into multidimensional
matrices. In some cases, like GIS and 3D modelling, the data are actually
points that lie in a two- or three-dimensional discretized space.
There are, however, other domains such as OLAP systems
\cite{OLAP1,OLAP2}
where the data are sets of tuples that are regarded as
entries in a multidimensional cube, with one dimension per attribute.
The domains of those attributes are not necessarily numeric, but may have richer
semantics. A typical case in OLAP
\cite{MULTIDIM1}, in particular in
snowflake schemes \cite{SNOWFLAKE},
is that each tuple contains a numeric summary
(e.g., amount of sales), which is regarded as the value of a cell in the data
cube. The domain of each dimension is hierarchical, so that each value in the
dimension corresponds to a leaf in a hierarchy (e.g., countries, cities,
and branches in one dimension, and years, months, and days in another).
Queries ask for summaries (sums, maxima, etc.) of all the cells that
are below some node of the hierarchy across each dimension (e.g., total sales
in New York during the previous month).

A way to handle OLAP data cubes is to linearize the hierarchy of the domain
of each dimension, so that each internal node corresponds to a range.
Summarization queries are then transformed into multidimensional range
queries, which are solved with multidimensional indexes \cite{Samet06}.
Such a structure is, however, more powerful than necessary, because it is
able to handle {\em any} multidimensional range, whereas the OLAP application
will only be interested in queries corresponding to combinations of nodes of
the hierarchies. There are well-known cases, in one dimension, of problems
that are more difficult for general ranges than if the possible
queries form a hierarchy. For example, categorical range counting queries
(i.e., count the number of different values in a range) requires in general
$\Omega(\log n / \log\log n)$ time if using $O(n\,\textrm{polylog}\, n)$
space \cite{LW13}, where $n$ is the array size, but if queries form a hierarchy
it is easily solved in constant time and $O(n)$ bits \cite{Sad07}. A second
example is the range mode problem (i.e., find the most frequent value in a
range), which is believed to require time $\Omega(n^{1.188})$ if using
$O(n^{1.188})$ space \cite{CDLMW12}, but if queries form a hierarchy it is
easily solved in constant time and linear space \cite{HSTV14}.

In this paper we aim at a compact data structure to represent data
cubes where the domains in each dimension are hierarchical.
Following the general idea of the tailored solutions to the problems
we mentioned \cite{Sad07,HSTV14}, our structure partitions the space
according to the hierarchies, instead of performing a regular
partition like generic multidimensional structures. Therefore, the
queries of interest for OLAP applications, which combine nodes of
the different hierarchies, will require aggregating the information
of just a few nodes in our partitions, much fewer than if we used a
generic space partitioning method.

Since we aim at compact representations, our baseline will be an
extension to multiple dimensions of a two-dimensional compact
summarization structure known as $k^2$-treap
\cite{k2treap:infosis2016}, a $k^2$-tree \cite{k2tree:infosis2014}
enriched with summary information on the internal nodes. This
$n$-dimensional treap, called $k^n$-treap, will then be extended so
that it can follow an arbitrary hierarchy, not only a regular one.
The topology of each hierarchy will be represented using a compact
tree representation, precisely LOUDS \cite{louds:Jacobson:1989}.
This new structure is called CMHD (Compact representation of
Multidimensional data on Hierarchical Domains). Although we focus on
sum queries in this paper, it is easy to extend our results to other
kinds of aggregations.

The rest of this paper is organized as follows.
Sections~\ref{sec:kntreaps} and \ref{sec:cmhd} describe our compact
baseline and then how it is extended to obtain our new data structure.
An experimental evaluation is given in Section~\ref{sec:experiments}.
Finally, we offer some conclusions and guidelines for future work.

\section{Our Baseline: $k^n$-treaps}\label{sec:kntreaps}

\begin{figure}[t]
\begin{center}
{\includegraphics[width=0.85\textwidth]{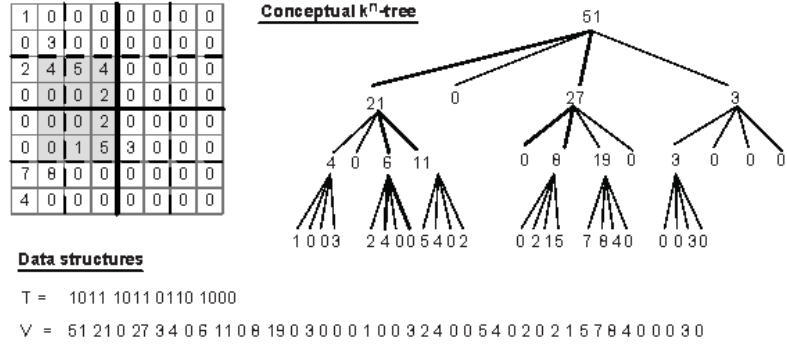}}
\end{center}
\vspace{-0.3cm} \caption{$k^n$-treap with a highlighted range query} \label{fig:kntreap}
\end{figure}

The $k^n$-treap is a straightforward extension of the $k^2$-treap to
manage multiple dimensions. It uses a $k^n$-tree (in turn a
straightforward extension of the $k^2$-tree) to store its topology,
and stores separately the list of aggregate values obtained from the
sum of all values in the corresponding submatrix.
Figure~\ref{fig:kntreap} shows a matrix and the corresponding
$k^n$-treap. The example uses two dimensions, but the same
algorithms are used for more dimensions.

Consider a hypercube of $n$ dimensions, where the length of each
dimension is $len = k^i$ for some $i$. If the length of the
dimensions are different, we can artificially extend the hypercube
with empty cells, with a minimum impact in the $k^n$-treap size. The
$k^n$-trees, which will be used to represent the $k^n$-treap
topology, are very efficient to represent wide empty areas. The
algorithm to build the $k^n$-treap starts storing on its root level
the sum of all values on the matrix%
\footnote{The implemented algorithm is recursive and each sum is
actually
computed only once, when returning from the recursive calls.}%
. It also splits each dimension into $k$ equal-sized parts, thus
giving a total of $k^n$ submatrices.
We define an ordering to traverse all the submatrices (in the example,
rows left-to-right, columns top-to-bottom).
Following this ordering,
we add a child node to the root for each submatrix.
The algorithm works recursively for each child node that represents a
nonempty submatrix, storing the sum of
the cells in this submatrix, splitting it and adding child nodes.
For empty sumatrices, the node stores a sum of $0$.

As we can see in Figure~\ref{fig:kntreap}, the root node stores
51, the sum of all values in the matrix, and it is decomposed into 4
matrices of size $4\times 4$, thus adding 4 children to the root node.
Notice that the second submatrix (top-right) is full of zeroes, so
this node just stores a sum of $0$ and is not further decomposed.
The algorithm proceeds recursively
for the remaining 3 children of the root node.

The final data structures used to represent the $k^n$-treap are the
following:
\begin{itemize}
\item \emph{Values (V)}: Contains the aggregated values (sums) for each
    (sub)matrix, as they would be obtained by a levelwise traversal of
    the $k^n$-treap. It is encoded using DACs \cite{dacs:ipm:2013}, which
    compress small values while allowing direct access.

\item \emph{Tree structure (T)}: It is a $k^n$-tree that stores a
    bitmap $T$ for the whole tree except its leaves. In this case,
    the usual bitmap $L$ for the leaves in a
    standard $k^n$-tree is not used, because  the information about which cells
    have or not a value is already represented in $V$. Therefore $L$  is not needed.
\end{itemize}

The navigation through the $k^n$-treap is basically a depth first
traversal. Finding the child of a node can be done very efficiently
by using $rank$ and $select$ operations \cite{louds:Jacobson:1989}
as in the standard $k^2$-tree. The typical queries in this context
are: finding the value of an individual cell and finding the sum of
the values in a given range of cells, specified by the initial and
final coordinates that define the submatrix of interest.
\paragraph{\bf{\emph{Finding the value of a specific cell by its coordinates.}}}
To find the value of the cell, for example the cell at coordinates
$(4,3)$ in the figure, the search starts at the root node and in
each step goes down trough the children of the matrix overlapping
the searched cell. In this example, the search goes through the
first child node (with value 21 in the figure), then through its
third child (with value 6) and finally through the second child,
reaching the leaf node with value 4, which is the value returned by
the query. 
\paragraph{\bf{\emph{Finding the sum of the cells in a submatrix.}}} The second
type of query looks for the aggregated value of a range of cells,
like the shaded area in Figure~\ref{fig:kntreap}. This is
implemented as a depth-first multi-branch traversal of the tree. If
the algorithm finds that the range specified in the query fully
contains a submatrix of the $k^n$-treap that has a precomputed sum,
it will use this sum and will not descend to its child nodes. The
figure highlights the branches of the $k^n$-treap that are used.
Notice that this query completely includes the sumatrices of values
$\{5,4,0,2\}$ and $\{0,2,1,5\}$, that have their sums (11 and 8)
explicitly stored on the third level of the tree. Therefore, the
algorithm does not need to reach the leaf levels of the tree for
these matrices. Notice also that there is an empty submatrix that
intersects with the region of the query (the first child of the 
third child of the root), so the algorithm also stops before
reaching the leaf levels in this submatrix. Only for cells $(3,2)$
(with a value of 4) and $(4,2)$ (with  a value of 0) needs the
algorithm to reach the leaf levels.
%

\section{Our proposal: CMHD}\label{sec:cmhd}

As previously stated, CMHD divides the matrix following the natural
hierarchy of the elements in each dimension. In this way we allow
the efficient answer of queries that consider the semantic of the
dimensions.

\subsection{Conceptual description}

Consider an $n$-dimensional matrix where each cell contains a weight
(e.g., product sales, credit card movements, ad views, etc.). The
CMHD recursively divides the matrix into several submatrices, taking
into account the limits imposed by the hierarchy levels of each
dimension.

\begin{figure}[t]
\begin{center}
{\includegraphics[width=1.00\textwidth]{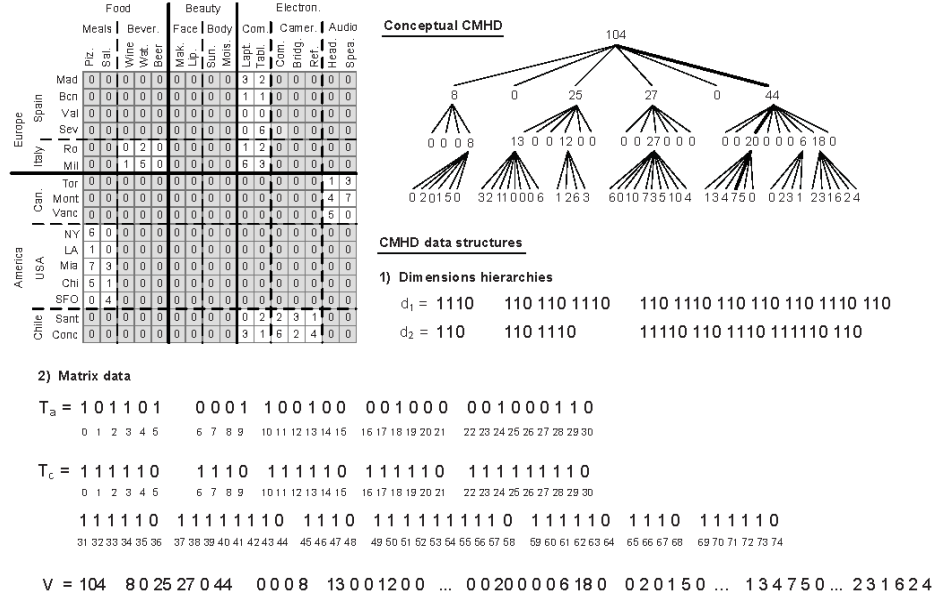}}
\end{center}
\vspace{-0.3cm} \caption{Example of CMHD construction for a
two-dimensional matrix.} \label{fig:cmhd}
\end{figure}

Figure \ref{fig:cmhd} depicts an example of a CMHD representation
for two dimensions. The matrix records the number of product sales
in different locations. For each dimension, a hierarchy of three
levels is considered. In particular, cities are aggregated into
countries and continents, while products are grouped into sections
and good categories. The tree at the right side of the image shows
the resulting conceptual CMHD for that matrix. Observe that each
hierarchy level leads to an irregular partition of the grid into
submatrices (each of them defined by the limits of its elements),
having as associated value the sum of product sales of the
individual cells inside it. Thus, the root of the tree stores the
total amount of sales in the complete matrix. Then the matrix is
subdivided by considering the partition corresponding to the first
level of the dimension hierarchies (see the bold lines). Each of the
submatrices will become a child node of the root, keeping the sum of
values of the cells in the corresponding submatrix. The
decomposition procedure is repeated for each child, considering
subsequent levels of the hierarchies (see the dotted lines), as
explained, until reaching the last one. Also notice that, as happens
in the $k^n$-treap, the decomposition concludes in all branches when
empty submatrices are reached (that is, in this scenario, when a
submatrix with no sales is found). See, for example, the second
child of the root.

Note that CMHD assumes the same height in all the hierarchies that
correspond to the different dimensions. Observe that, for each
crossing of elements of the same level from different dimensions, an
aggregate value is stored. Notice also that artificial levels can be
easily added to a hierarchy of one dimension by subdividing all the
elements of a level in just one element (itself), thus creating a
new level identical to the previous one.  This feature allows us to
arbitrarily match the levels of the different hierarchies, and thus
to flexibly adapt the representation of aggregated data to
particular query needs. That is, by introducing artificial
intermediate levels where required, more interesting aggregated
values will be precomputed and stored. For example, assume we have
two dimensions: ($d_{1}$) with levels for \emph{department},
\emph{section} and \emph{product}; and ($d_{2}$) with levels for
\emph{year}, \emph{season}, \emph{month} and \emph{day}. If we were
interested in obtaining the number of sales per \emph{section} for
\emph{season}s, but also for \emph{months}, we could devise a new
level arrangement for $d_{1}$, that will have now the levels
\emph{department}, \emph{section}, \emph{section'}, \emph{product};
where each particular \emph{section} of the second hierarchy level
results into just one \emph{section'} child, which is actually
itself. In this way aggregated values will be computed and stored
considering sales for \emph{section} in each \emph{season}, but also
sales for \emph{section'} in each \emph{month}. 

\subsection{Data structures}

The conceptual tree that defines the CMHD is represented compactly
with different data structures, for the domain hierarchies and for
the matrix data itself. 

\paragraph{\bf{\emph{Domain hierarchy representation.}}} The hierarchy of a
dimension domain
is essentially a tree of $C$ nodes. We represent this tree using
LOUDS \cite{louds:Jacobson:1989}, a tree representation that uses $2C$
bits, and can efficiently navigate it.
Using LOUDS, a tree representing  the hierarchy
of the elements of a dimension is encoded by appending the degree $r$
of each node in (left-to-right) level-order, in unary:
$1^{r}0$. Figure \ref{fig:cmhd} illustrates the hierarchy encoding
of the dimensions used in that example (see $d_{1}$ and $d_{2}$).
For instance, the degree of the first node for the products
dimension ($d_{1}$) is 3, so its unary encoding is 1110. Note that
each node (i.e., element of a dimension placed at any level of its
hierarchy) is associated with one 1 in the encoded representation of
the degree of its parent. LOUDS is navigated using $rank$ and $select$
queries: $rank_b(i)$ is the number of bits $b$ up to position $i$, and
$select_b(j)$ is the position of the $j$th occurrence of bit $b$. Both
queries are computed in constant time using $o(C)$ additional bits \cite{Cla96}.
For example, given a node whose unary representation starts at position $i$,
its parent is $p=select_0(t-j)+1$, where $t=select_1(j)$ and $j=rank_0(i)$; and
$i$ is the $(t-p+1)$th child of $p$. On the other hand, the $k$th child of $i$
is $select_0(rank_1(i)+k-1)+1$.
We also use a hash table to associate the domain nodes (i.e., labels such as
``USA'' in Figure \ref{fig:cmhd}) with the corresponding LOUDS node position.

\paragraph{\bf{\emph{Data representation.}}} To represent the $n$-dimensional matrix,
we use the following data structures:

\begin{itemize}

\item \emph{Tree structures ($T_{a}$ and $T_{c}$)}: to navigate the
CMHD, we need to use two different data structures in conjunction.
First, $T_{a}$, a bit array that, similarly to the $k^{n}$-treap,
provides a compact representation of the conceptual tree
independently of the node values, for all the tree levels, except
the last one\footnote{We do not actually need to represent the nodes
of the last level in $T_{a}$. This data structure will be used to
first identify a node whose children will be later located in
another bit array ($T_{c}$). But these already constitute matrix
cells, with no children.}. That is, internal nodes whose associated
value is greater than 0, will be represented with a 1. In other
case, they will be labeled with a 0. Observe that, for the
$k^{n}$-treap, the use of this data structure is enough to navigate
the tree, taking advantage of the regular partition of the matrix
into equal-sized submatrices. Instead, CMHD follows different
hierarchy partitions, which results into irregular submatrices.
Therefore, a second data structure, $T_{c}$, is also required to
traverse the CMHD. This is a bit array aligned to $T_{a}$, which
marks the limits of each tree node in $T_{a}$ (this time, it also
considers the last tree level). If the next tree node in $T_a$ has
$z$ children, we append $1^{z-1}0$ to $T_c$. Notice that each node
of $T_a$ is associated with a 0 in $T_c$, which allows navigating
the trees using $rank$ and $select$ on $T_a$ and $T_c$: say we are
at a node in $T_a$ that starts at position $i$; then it has a $k$th
child iff $T_a[i+k-1]=1$, and if so this child starts at position
$select_0(T_c,rank_1(T_a,i+k-1))+1$.


\item \emph{Values (V)}: the CMHD is traversed levelwise
storing the values associated with each node 
(either corresponding to original matrix cells, or to
data aggregations) in a single sequence, which is then represented
with DACs \cite{dacs:ipm:2013}.
\end{itemize}

\subsection{Queries}

Queries in this context give the names of elements of the different
dimensions and ask for the sum of the cells defined for those
values. Depending on the query, we can answer it by just reporting a
single aggregated value already kept in \emph{V}, or by retrieving
several stored values, and then adding them up. The first scenario
arises when the elements (labels) of the different dimensions
specified in the query are all at the same level in their respective
hierarchies. The second situation arises from queries using labels
of different levels. In both contexts, top-down traversals of the
conceptual CMHD are required to fetch the values.
The algorithm always starts searching the hash tables for the labels provided
by the query for the different dimensions, to locate the corresponding
LOUDS nodes. From the LOUDS nodes, we traverse each hierarchy upwards to
find out its depth and the child that must be followed at each level to reach
it.

This information is then used to find the desired nodes in $T_a$. For example,
with two dimensions, we start at the root of $T_a$ and descend to the child
number $k_1 + a_1 \cdot k_2$, where $k_i$ is the child that must be followed
in the $i$th dimension to reach the queried node, and $a_i$ is the number of
children of the root in the $i$th dimension ($a_i$ is easily computed with
the LOUDS tree of its dimension). We continue similarly to the node at level 2,
and so on, until we reach one of the query nodes in a dimension,
say in the first. Now, to reach the other (deeper) node in the second dimension,
we must descend by every child in the first dimension, at every level, until
reaching the second queried node.
Finally, when we have reached all the nodes, we collect and sum up
the corresponding values from $V$.
Note that, if all the queried nodes are in the same level, we perform a
single traversal in $T_a$.
Note also that, if we find any zero in a node of $T_a$ along this traversal,
we immediately prune that branch, as the submatrix contains no data.

\paragraph{Example.} Assume we want to retrieve the total
amount of \emph{speaker} sales in \emph{Montreal}, in Figure
\ref{fig:cmhd}. Since both labels belong to the same level in both
dimension hierarchies (the last one), we will have to retrieve a
single stored value in that level. The path to reach it has been
highlighted in the conceptual tree of the image. To perform the
navigation we must start at the root of the tree (position 0 in
$T_{a}$). In the first level, we need to fetch the sixth child
(offset 5), as it corresponds to the submatrix including the element
to search, in that level. Hence we access position 5 in $T_{a}$.
Since $T_{a}[5] = 1$, we must continue descending to the next level.
Recall that we have a 1 in $T_{a}$ for each node with children, and
that each node is associated with just one 0 in $T_{c}$. So the
child starts at position $select_{0}(T_{c}, rank_{1}(T_{a},5))+1 =
select_{0}(T_{c}, 4)+1 = 22$ in $T_{a}$. In this level we must
access the third child (offset 2), so we check $T_{a}[24] = 1$.
Again, as we are in an internal node, we know that its children are
located at position $select_{0}(T_{c}, rank_{1}(T_{a}, 24)) + 1 =
select_{0}(T_{c}, 9) + 1 = 59$. Finally, we reach the third and last
level of the tree, where we know that the corresponding child is the
fourth one (at $T_a[59+3]=T_a[62]$). Recall, however, that this last
level is not represented in $T_{a}$. To perform this final step, we
directly look into the array $V$: $V[62+1] = V[63] = 7$ is the
answer.

In case of queries combining labels of different levels, the same
procedure would apply, but having to get the values corresponding to
all the possible combinations with the element of the lowest
hierarchy level (e.g., if we want to obtain the number of \emph{meal}
sales in \emph{America}, we must first recover the values associated
with \emph{meal}-\emph{Canada}, \emph{meal}-\emph{USA}, and
\emph{meal}-\emph{Chile}, and then sum them up).

\section{Experimental Evaluation}\label{sec:experiments}
This section presents the empirical evaluation of the two previously
described data structures. Both representations have been
implemented in C/C++, and the compiler used was GCC 4.6.1. (option
-O$9$). We ran our experiments in a dedicated Intel(R) Core(TM)
i7-3820 CPU @ 3.60GHz (4 cores) with 10MB of cache, and 64GB of RAM.
The machine runs Ubuntu 12.04.5 LTS with kernel 3.2.0-99 (64 bits).

We generate different datasets, all of them synthetic, to evaluate
the performance of the two data structures, varying the number of
dimensions and the number of items on each dimension. These datasets
have been labeled as \verb!<dim#>D_<item#>!, thus referring to their
size specifications in the own name. For example, dataset
\verb!5D_16! has 5 dimensions, and the number of items on each
dimension is 16. The total size of this dataset is $16^5=1048576$
elements.

In order to show the CMHD advantage of considering the domain
semantics, and computing the aggregate values according to the
natural limits imposed by the hierarchy of elements in each
dimension, the dimensions hierarchies have been generated in two
different ways for each dataset. First, the \emph{binary}
organization, that corresponds to a regular partition. That is, the
hierarchies of each dimension are exactly the same as those produced
by a $k^n$-treap matrix partition into equal-sized submatrices. In
this way both data structures store exactly the same aggregated
values. We named it \emph{binary} because we use a value of $k=2$. 
Second, the \emph{irregular} organization, which
arbitrary groups data, on each dimension, into different and
irregular hierarchies (different number of divisions, and also
different size at each level). The last scenario simulates what
would be a matrix partition following the semantic needs of a given
domain. In this case the aggregated values stored by the CMHD will
be different from those stored by the $k^n$-treap, and therefore
more appropriated to answer queries using the same ``semantic''.
That means, in our context, queries considering regions that exactly
match the natural divisions of each dimension at some level of the
hierarchies.

To test the structures behavior, we have also considered three
different datasets, with a different number of empty cells, for each
size specification: with no empty cells, and with 25\% and 50\% of
empty cells, respectively.

First we analyze the space requirements of both data structures for
all the datasets (see Table \ref{table:sizes}). Of course, the size
decreases as the number of empty cells increases, in both cases.
Moreover, we can also observe that the $k^n$-treap size is slightly
lower than the CMHD. This is expected, because CMHD has to store the
LOUDS representation of each dimension hierarchy, while dimensions
are implicit for the $k^n$-treap. Additionally, CMHD uses a second
bitmap ($T_c$) to navigate the conceptual tree, which is not
necessary when using the $k^n$-treap.

We must also clarify a small issue about the sizes of the
$k^n$-treaps: the size of a standard $k^n$-treap for a specific
dataset is always the same, regardless of the organization of its
dimensions (binary or irregular). However, Table \ref{table:sizes}
shows some difference in the sizes. 
For example, for \verb!4D_16!, the size
for the binary organization is $44.84$, but it is $44.42$ for the
irregular one. The reason for this variation is that all queries are
performed by taking dimension labels as input, so we need a
vocabulary to translate each label into a range of cells. We have
included that vocabulary (dimension labels and cell ranges) into the
size of the $k^n$-treaps, and the vocabulary for the irregular
organization is usually smaller, as it has less levels and less
dimension labels (because each node in the conceptual tree can have
more than 2 children in the irregular organization, while the binary
organization always has 2).

\begin{table}[th]
    \centering
    \scalebox{0.7}{
    \begin{tabular}{l|r|r|r|r|r|r|r|r|r|r|r|r|}
        \cline{2-13}
             & \multicolumn{4}{c|}{\textbf{0\% Zeroes}} & \multicolumn{4}{c|}{\textbf{25\% Zeroes}} & \multicolumn{4}{c|}{\textbf{50\% Zeroes}} \\
        \cline{2-13}
             & \multicolumn{2}{|c|}{\textbf{Binary}} & \multicolumn{2}{c|}{\textbf{Irregular}} & \multicolumn{2}{c|}{\textbf{Binary}}
             & \multicolumn{2}{c|}{\textbf{Irregular}} & \multicolumn{2}{c|}{\textbf{Binary}} & \multicolumn{2}{c|}{\textbf{Irregular}} \\
        \hline
        \multicolumn{1}{|l|}{\textbf{name}} & \textbf{kn-treap} & \textbf{CMHD} & \textbf{kn-treap} & \textbf{CMHD} & \textbf{kn-treap} & \textbf{CMHD} & \textbf{kn-treap} & \textbf{CMHD} & \textbf{kn-treap} & \textbf{CMHD} & \textbf{kn-treap} & \textbf{CMHD}  \\
        \hline
\multicolumn{1}{|l|}{\texttt{4D\_16}} &  44.84 & 55.56 & 44.22 & 47.82 & 38.16 & 47.54 & 37.54 & 43.04 & 29.51 & 37.09 & 28.89 & 34.18  \\
        \hline
\multicolumn{1}{|l|}{\texttt{4D\_32}} & 680.45 & 864.17 & 679.08 & 750.17 & 552.15 & 710.05 & 550.78 & 640.80 & 400.13 & 527.57 & 398.76 & 501.01  \\
        \hline
\multicolumn{1}{|l|}{\texttt{5D\_16}} & 631.10 & 793.34 & 630.41 & 729.48 & 527.23 & 667.69 & 526.54 & 653.25 & 408.10 & 523.27 & 407.41 & 509.80  \\
        \hline
\multicolumn{1}{|l|}{\texttt{5D\_32}} & 20098.99 & 25328.26 & 20097.43 & 23344.18 & 16272.61 & 20691.87 & 16271.04 & 20167.62 & 11776.94 & 15237.57 & 11775.37 & 15471.61  \\
        \hline
\multicolumn{1}{|l|}{\texttt{6D\_16}} & 9663.37 & 12073.82 & 9662.48 & 11456.36 & 8180.31 & 10278.04 & 8179.43 & 10532.61 & 6419.99 & 8135.86 & 6419.11 & 8279.60 \\
        \hline

\end{tabular}
}
    \caption{Space requirements of $k^n$-treap and CMHD data structures (in KB)}\label{table:sizes}

\end{table}

Regarding query times, we have run several sets of queries for all
the datasets. As previously mentioned, queries are posed in this
context by giving one element name (label) for each different
dimension, as it is the natural way to query a multidimensional
matrix defined by hierarchical dimensions. Since the $k^n$-treap
does not directly work with labels, each query has been translated
into the equivalent ranged query, establishing the initial and final
coordinates for each dimension.
The following types of queries have been considered:

\begin{itemize}
\item \emph{Finding one precomputed value}. This value can be
    a specific cell of the matrix (so forcing the algorithms to reach the last
    level of the tree), or a precomputed value that corresponds
    to an internal node of the conceptual tree.

    Following the example of Figure~\ref{fig:cmhd}, a query asking
    for the amount of \emph{speakers} sales in \emph{Montreal} or the total number of  \emph{beberages} sales in
    \emph{Italy} would be queries of this type, the former accessing an individual
    cell and the later obtaining a precomputed value in the penultimate
    level of the tree.
\vspace{2mm}
\item \emph{Finding the sum of several precomputed values}.
    This kind of query must obtain a sum that is not precomputed and stored in
    the data structure itself. In turn, it must access several
    of these aggregated values and then add them up. Given that we are specifying the queries by dimension labels,
    this type of query is defined by using labels that belong to different
    levels of the hierarchies across the dimensions. The lowest level,
    which corresponds to individual cells, is not used for this scenario.

    An example of this query type would be to find the total number
    of sales of \emph{electronic} products in Chile. Note that \emph{electronic} is located
    at the first level of its dimension hierarchy, but \emph{Chile} is at the second level
    of the second dimension (see Figure~\ref{fig:cmhd}). Hence, the values corresponding to
    \emph{computers-Chile}, \emph{cameras-Chile}, and \emph{audio-Chile} must be first retrieved
    to finally sum them up.
\end{itemize}

Each created set contains $10,000$ queries, randomly generated, of the
two previous types, for each dataset. The following tables show the
average query times (in microseconds per query) for both data
structures, taking into account the two different matrix partitions
of the datasets (\emph{binary} or \emph{irregular}) and also the
percentage of empty cells.

We first show the results obtained for queries that just need to
retrieve one precomputed value, at different levels. On the one
hand, Table~\ref{table:queries1} displays query times for specific
matrix cells, that is, located at the last level of the conceptual
tree. In this case, the $k^n$-treap performs better than the CMHD in
almost all cases. This is an expected outcome as both data
structures must reach the leaf level to get the answer, and the
depth first navigation of the tree is simpler in the 
$k^n$-treap (just products and $rank$ operations). In any case, CMHD
also performs quite well, using just a few microseconds to answer
any of the queries.

\begin{table}[!htb]
    \centering
    \scalebox{0.7}{
    \begin{tabular}{l|r|r|r|r|r|r|r|r|r|r|r|r|}
        \cline{2-13}
             & \multicolumn{4}{c|}{\textbf{0\% Zeroes}} & \multicolumn{4}{c|}{\textbf{25\% Zeroes}} & \multicolumn{4}{c|}{\textbf{50\% Zeroes}} \\
        \cline{2-13}
             & \multicolumn{2}{|c|}{\textbf{Binary}} & \multicolumn{2}{c|}{\textbf{Irregular}} & \multicolumn{2}{c|}{\textbf{Binary}}
             & \multicolumn{2}{c|}{\textbf{Irregular}} & \multicolumn{2}{c|}{\textbf{Binary}} & \multicolumn{2}{c|}{\textbf{Irregular}} \\
        \hline
        \multicolumn{1}{|l|}{\textbf{Dataset}} & \textbf{kn-treap} & \textbf{CMHD} & \textbf{kn-treap} & \textbf{CMHD} & \textbf{kn-treap} & \textbf{CMHD} & \textbf{kn-treap} & \textbf{CMHD} & \textbf{kn-treap} & \textbf{CMHD} & \textbf{kn-treap} & \textbf{CMHD}  \\
        \hline
        \multicolumn{1}{|l|}{\texttt{4D\_16}} & 2 & 4 & 2 & 3 & 2 & 4 & 2 & 2 & 2 & 4 & 2 & 3 \\
        \hline
        \multicolumn{1}{|l|}{\texttt{4D\_32}} & 2 & 5 & 3 & 4 & 2 & 4 & 2 & 1 & 2 & 4 & 1 & 4 \\
        \hline
        \multicolumn{1}{|l|}{\texttt{5D\_16}} & 2 & 4 & 3 & 4 & 2 & 5 & 3 & 2 & 2 & 5 & 2 & 2 \\
        \hline
        \multicolumn{1}{|l|}{\texttt{5D\_32}} & 3 & 6 & 3 & 5 & 2 & 4 & 3 & 3 & 2 & 6 & 3 & 2 \\
        \hline
        \multicolumn{1}{|l|}{\texttt{6D\_16}} & 3 & 4 & 3 & 4 & 3 & 6 & 4 & 2 & 4 & 5 & 4 & 4 \\
        \hline
\end{tabular}
}
    \caption{Average query times (in $\mu$s) for queries finding one precomputed value (original matrix cells).}\label{table:queries1}
\end{table}

On the other hand, Table~\ref{table:queries2} shows the average
query times for queries of the same type, but now considering
precomputed values stored in nodes of an intermediate level of the
tree (in particular, the penultimate level). Note that this fact
holds for both data structures when working with a regular partition
of the matrix (that is, the \emph{binary} scenario). Thus, in this
case, the $k^n$-treap gets better results than CMHD, but with slight
time differences. Yet, observe that this is not the actual scenario
when dealing with meaningful application domains, where rich
semantics arise. This situation is that corresponding to what we
called \emph{irregular} datasets. In this case, CMHD excels, as
expected, given that this data structure has been particularly
designed to manage hierarchical domains. Results show that CMHD is
able to perform up to 12 times faster than $k^n$-treap (for the best
case).



\begin{table}[!htb]
    \centering
    \scalebox{0.7}{
    \begin{tabular}{l|r|r|r|r|r|r|r|r|r|r|r|r|}
        \cline{2-13}
             & \multicolumn{4}{c|}{\textbf{0\% Zeroes}} & \multicolumn{4}{c|}{\textbf{25\% Zeroes}} & \multicolumn{4}{c|}{\textbf{50\% Zeroes}} \\
        \cline{2-13}
             & \multicolumn{2}{|c|}{\textbf{Binary}} & \multicolumn{2}{c|}{\textbf{Irregular}} & \multicolumn{2}{c|}{\textbf{Binary}}
             & \multicolumn{2}{c|}{\textbf{Irregular}} & \multicolumn{2}{c|}{\textbf{Binary}} & \multicolumn{2}{c|}{\textbf{Irregular}} \\
        \hline
        \multicolumn{1}{|l|}{\textbf{Dataset}} & \textbf{kn-treap} & \textbf{CMHD} & \textbf{kn-treap} & \textbf{CMHD} & \textbf{kn-treap} & \textbf{CMHD} & \textbf{kn-treap} & \textbf{CMHD} & \textbf{kn-treap} & \textbf{CMHD} & \textbf{kn-treap} & \textbf{CMHD}  \\
        \hline

        \multicolumn{1}{|l|}{\texttt{4D\_16}} & 1 & 4 & 7 & 3 & 1 & 3 & 6    & 2 & 2 & 4 & 5 & 2 \\
       \hline
        \multicolumn{1}{|l|}{\texttt{4D\_32}} & 1 & 3 & 9 & 3 & 1 & 3 & 7 & 2 & 1 & 3 & 6 & 1 \\
       \hline
        \multicolumn{1}{|l|}{\texttt{5D\_16}} & 2 & 4 &   11 & 1 & 2 & 3   & 9 & 1 & 2 & 4 & 7 & 3 \\
      \hline
        \multicolumn{1}{|l|}{\texttt{5D\_32}} & 3 & 4 & 23 & 2 & 2 & 4 & 18 & 3 & 2 & 3 & 12 & 2 \\
       \hline
       \multicolumn{1}{|l|}{\texttt{6D\_16}} & 2 & 4 & 35 & 2 & 2 & 2 & 28 & 3 & 3 & 4 & 21 & 1 \\
       \hline
\end{tabular}
}
    \caption{Average query times (in $\mu$s) for queries finding one precomputed value (penultimate tree level).}\label{table:queries2}
\end{table}


To check whether the observed differences are significative (in the
cases where times were closer) we performed a statistical
significance test. We checked the \verb!4D_16! and \verb!5D_16!
datasets, for the irregular organization, with all the different
configurations of empty cells.

We show here, as a proof, the details for \verb!4D_16! with 50\% of
empty cells, which took $5 \mu\mbox{s}$ to the $k^n$-treap, and $2
\mu\mbox{s}$ to CMHD. We ran 20 sets of $10,000$ queries, and
measured both the average time and the standard deviation for the
$k^n$-treap ($5.100$ and $0.447$, respectively) and for the CMHD
($1.750$ and $0.550$, respectively). With these results, we obtain a
critical value of $4.725$, which is greater than $2.580$, so the
difference is significative with a 99\% of confidence level. The
remaining tests also proved the same significance results.

Finally, Table~\ref{table:queries4} presents the average query times
for the second type of queries (that is, those having to recover
several precomputed values and then adding them up to provide the
final answer). As results show, the $k^n$-treap displays a better
performance than CMHD for the \emph{binary} scenario. However, again
this is not the most interesting situation in real domains. If we
observe the results obtained for the \emph{irregular} datasets, we
will appreciate that CMHD clearly outperforms the $k^n$-treap in
this scenario, thus demonstrating the good capabilities of our
proposal to cope with the aim of this work.

\begin{table}[!htb]
    \centering
    \scalebox{0.7}{
    \begin{tabular}{l|r|r|r|r|r|r|r|r|r|r|r|r|}
        \cline{2-13}
             & \multicolumn{4}{c|}{\textbf{0\% Zeroes}} & \multicolumn{4}{c|}{\textbf{25\% Zeroes}} & \multicolumn{4}{c|}{\textbf{50\% Zeroes}} \\
        \cline{2-13}
             & \multicolumn{2}{|c|}{\textbf{Binary}} & \multicolumn{2}{c|}{\textbf{Irregular}} & \multicolumn{2}{c|}{\textbf{Binary}}
             & \multicolumn{2}{c|}{\textbf{Irregular}} & \multicolumn{2}{c|}{\textbf{Binary}} & \multicolumn{2}{c|}{\textbf{Irregular}} \\
        \hline
        \multicolumn{1}{|l|}{\textbf{Dataset}} & \textbf{kn-treap} & \textbf{CMHD} & \textbf{kn-treap} & \textbf{CMHD} & \textbf{kn-treap} & \textbf{CMHD} & \textbf{kn-treap} & \textbf{CMHD} & \textbf{kn-treap} & \textbf{CMHD} & \textbf{kn-treap} & \textbf{CMHD}  \\
        \hline

\multicolumn{1}{|l|}{\texttt{4D\_16}} & 3 & 10 & 20 & 4 & 3 & 6 & 16 & 3 & 2 & 6 & 12 & 6\\
       \hline
\multicolumn{1}{|l|}{\texttt{4D\_32}} & 6 & 21 & 21 & 3 & 4 & 20 & 17 & 4 & 4 & 20 & 12 & 5\\
       \hline
\multicolumn{1}{|l|}{\texttt{5D\_16}} & 4 & 8 & 30 & 8 & 4 & 8 & 25 & 3 & 3 & 7 & 19 & 1\\
       \hline
\multicolumn{1}{|l|}{\texttt{5D\_32}} & 6 & 26 & 49 & 3 & 8 & 23 & 39 & 5 & 6 & 21 & 27 & 2\\
       \hline
\multicolumn{1}{|l|}{\texttt{6D\_16}} & 5 & 15 & 106 & 7 & 5 & 10 & 82 & 6 & 4 & 9 & 63 & 8\\
       \hline
\end{tabular}
}
    \caption{Average query times (in $\mu$s) for queries finding a sum of precomputed values.}\label{table:queries4}
\end{table}

\section{Conclusions and Future Work}\label{sec:conclusions}

We have presented a multidimensional compact data structure that is
tailored to perform aggregate queries on data cubes over
hierarchical domains, rather than general range queries. The
structure represents each hierarchy with a succinct tree
representation, and then partitions the data cube according to the
product of the hierarchies. This partition is represented with an
extension of the $k^2$-treap to higher dimensions and to non-regular
partitions. The resulting structure, dubbed CMHD, is much faster
than a regular multidimensional $k^2$-treap when the queries follow
the hierarchical domains. This makes it particularly attractive to
represent OLAP data cubes compactly and efficiently answer
meaningful aggregate queries.

As future work, we plan to experiment on much larger collections.
This would make the vocabulary of hierarchy nodes much less
significant compared to the data itself (especially for the CMHD).
We also plan to test real datasets (for example, coming from data
warehouses) and real query workloads. We also expect to compare our
results with established OLAP database management systems, and to
enrich our prototype with other kinds of queries and data.

\bibliographystyle{splncs03}
\bibliography{paper}

\end{document}